\documentclass[%
aps,
 superscriptaddress,
 twocolumn,
 showpacs,
preprintnumbers,
nofootinbib,
 amsmath,amssymb,
 prd,
balancelastpage,
nofootinbib,
floatfix
]{revtex4-1}

\usepackage{graphicx}
\usepackage{dcolumn}
\usepackage{bm}
\usepackage{amsmath}
\usepackage{diagbox}
\usepackage{rotating}
\usepackage{multirow}
\usepackage{xcolor}
\usepackage{hepunits}
\usepackage{comment}

\usepackage{hyperref}
\usepackage[mathlines]{lineno}
\newcommand{\be}{ \begin{eqnarray} }
\newcommand{\ee}{ \end{eqnarray} }
\newcommand{\vecq}{\mathbf{q}}

\begin{document}


\preprint{FERMILAB-PUB-22-044-PPD-QIS-T}

\title{Revisiting the Dark Matter Interpretation of Excess Rates in Semiconductors}

\author{Peter~Abbamonte}\thanks{abbamont@illinois.edu}
\affiliation{Department of Physics, University of Illinois at Urbana-Champaign, Urbana, IL 61801, USA}

\author{Daniel~Baxter}\thanks{dbaxter9@fnal.gov}
\affiliation{Fermi National Accelerator Laboratory, Batavia, IL 60510, USA}
 
\author{Yonatan~Kahn}\thanks{yfkahn@illinois.edu}
\affiliation{Department of Physics, University of Illinois at Urbana-Champaign, Urbana, IL 61801, USA}
\affiliation{Illinois Center for Advanced Studies of the Universe, University of Illinois at Urbana-Champaign, Urbana, IL 61801, USA}

\author{Gordan~Krnjaic}\thanks{krnjaicg@fnal.gov}
\affiliation{Fermi National Accelerator Laboratory, Batavia, IL 60510, USA}
\affiliation{
Department of Astronomy and Astrophysics, University of Chicago,, Chicago, IL 60637, USA}
\affiliation{Kavli Institute for Cosmological Physics, University of Chicago, Chicago, IL 60637, USA}

\author{Noah~Kurinsky}\thanks{kurinsky@slac.stanford.edu}
\affiliation{SLAC National Accelerator Laboratory, 2575 Sand Hill Road, Menlo Park, CA 94025, USA}

\author{Bashi~Mandava}\thanks{bmandava@berkeley.edu}
\affiliation{Department of Physics, University of Illinois at Urbana-Champaign, Urbana, IL 61801, USA}
\affiliation{Berkeley Center for Theoretical Physics, University of California, Berkeley, CA 94720, USA}

\author{Lucas~K.~Wagner}\thanks{lkwagner@illinois.edu}
\affiliation{Department of Physics, University of Illinois at Urbana-Champaign, Urbana, IL 61801, USA}
\affiliation{Institute for Condensed Matter Theory, University of Illinois at Urbana-Champaign, Urbana, IL 61801, USA}

\date{\today}

\begin{abstract}
In light of recent results from low-threshold dark matter detectors, we revisit the possibility of a common dark matter origin for multiple excesses across numerous direct detection experiments, with a focus on the excess rates in semiconductor detectors. We explore the interpretation of the low-threshold calorimetric excess rates above 40~eV in the silicon SuperCDMS Cryogenic Phonon Detector and above 100~eV in the germanium EDELWEISS Surface detector as arising from a common but unknown origin, and demonstrate a compatible fit for the observed energy spectra in both experiments, which follow a power law of index $\alpha = 3.43^{+0.11}_{-0.06}$. Despite the intriguing scaling of the normalization of these two excess rates with approximately the square of the mass number $A^2$, we argue that the possibility of common origin by dark matter scattering via nuclear recoils is strongly disfavored, even allowing for exotic condensed matter effects in an as-yet unmeasured kinematic regime, due to the unphysically-large dark matter velocity required to give comparable rates in the different energy ranges of the silicon and germanium excesses. We also investigate the possibility of inelastic nuclear scattering by cosmic ray neutrons, solar neutrinos, and photons as the origin, and quantitatively disfavor all three based on known fluxes of particles. 
\end{abstract}

\maketitle

\section{Introduction} 
\label{sec:Intro}

Direct detection experiments searching for particle dark matter (DM) with masses below 1~GeV have made significant advancements in the last decade, driven by lower thresholds, improved resolution, and sophisticated analysis techniques~\cite{Proceedings:2022hmu}. 
These experiments are on the forefront of new technological development, and have demonstrated sensitivity to individual electron-hole pair creation at the eV energy scale~\cite{Crisler:2018gci,Abramoff_2019,barak2020sensei,Agnese_2018,Castello-Mor:2020jhd,edelweissHV} as well as eV-scale calorimetry enabling direct energy measurements independent of charge production~\cite{alkhatib2020light,EdelweissWIMP,Abdelhameed_2019}. An important distinction between ionization and calorimetric detectors is that ionization detectors are all limited by uncalibrated, non-radiogenic backgrounds which are often referred to as dark rates. A dark rate can in principle arise from any source that produces anomalous ionization events in a detector, with an irreducible contribution from thermal processes at the detector temperature. Substantial effort is under way to better characterize these dark rates~\cite{senseiDarkRate,Du:2020ldo}. 
On the other hand, calorimetric detectors currently have higher energy thresholds but do not suffer from the dark rates mentioned above. This complementarity offers an interesting window on new physics when the two detector types are taken together, as was previously done in Ref.~\cite{plasmonKBKK}.

In this paper, we continue in the spirit of Ref.~\cite{plasmonKBKK} by performing a joint analysis of the two most recent results from calorimetric semiconductor detectors, the silicon SuperCDMS Cryogenic Phonon Detector (SuperCDMS CPD)~\cite{alkhatib2020light} and the germanium EDELWEISS Surface detector (EDELWEISS-Surf)~\cite{EdelweissWIMP}. Both experiments observe a statistically significant excess event rate above known background sources near threshold. Our analysis here differs from Ref.~\cite{plasmonKBKK} because recent work has sharply constrained our previously-proposed signal models: the plasmon production channel from nuclear scattering is only a small part of the total spectrum from the Migdal effect in solid-state systems and cannot account for the observed spectral shape~\cite{Kozaczuk:2020uzb,Knapen:2020aky}, and a fast DM subcomponent is excluded by XENON1T except for a very narrow range of DM velocities~\cite{Harnik:2020ugb}. That said, our approach is similar in that we consider novel inelastic nuclear scattering channels where the relationship between the deposited energy $E_r$ and the momentum transfer from the DM $q$ differs from $E_r = q^2/(2m_N)$ (where $m_N$ is the mass of the nucleus) expected from free-particle elastic scattering.\footnote{Inelasticity here refers exclusively to detector response and is not to be confused with inelastic DM, which is a mass splitting between different DM states \cite{TuckerSmith:2001hy}.} Indeed, given that the energy scales of the excess are close to the lattice displacement energy, many-body effects may be expected to be important~\cite{Kahn:2020fef}, and collective effects do substantially extend the reach of semiconductor ionization detectors to sub-GeV DM through the Migdal effect~\cite{Knapen:2020aky}, compared to calculations which assume isolated atom targets~\cite{Ibe:2017yqa,Dolan:2017xbu,Bell:2019egg,migdalEssig,2020arXiv200710965L}.

\begin{figure*}[t]
    \centering
    \includegraphics[width=0.49\textwidth, trim=20 0 40 20, clip=true]{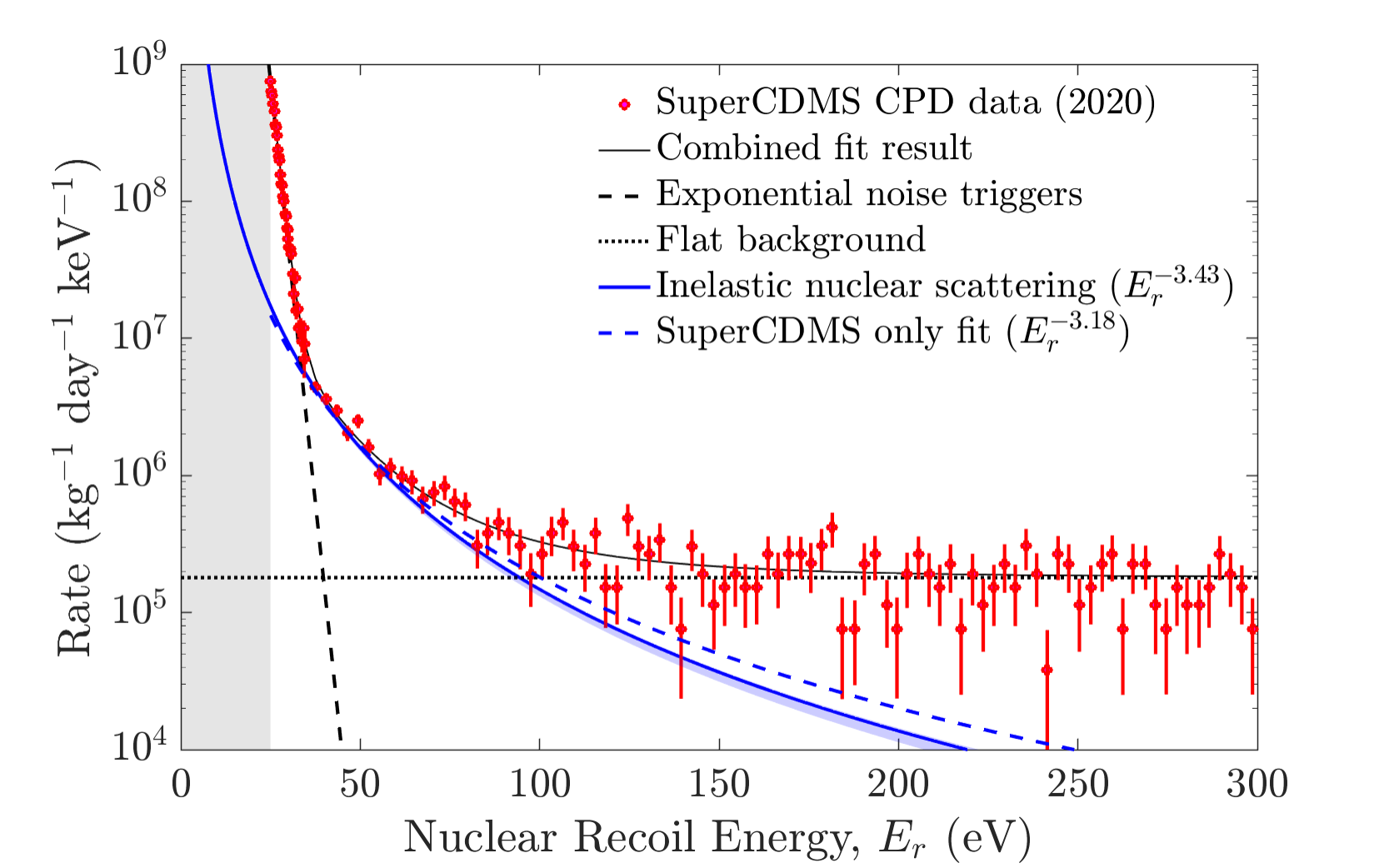}
    \includegraphics[width=0.49\textwidth, trim=20 0 40 20, clip=true]{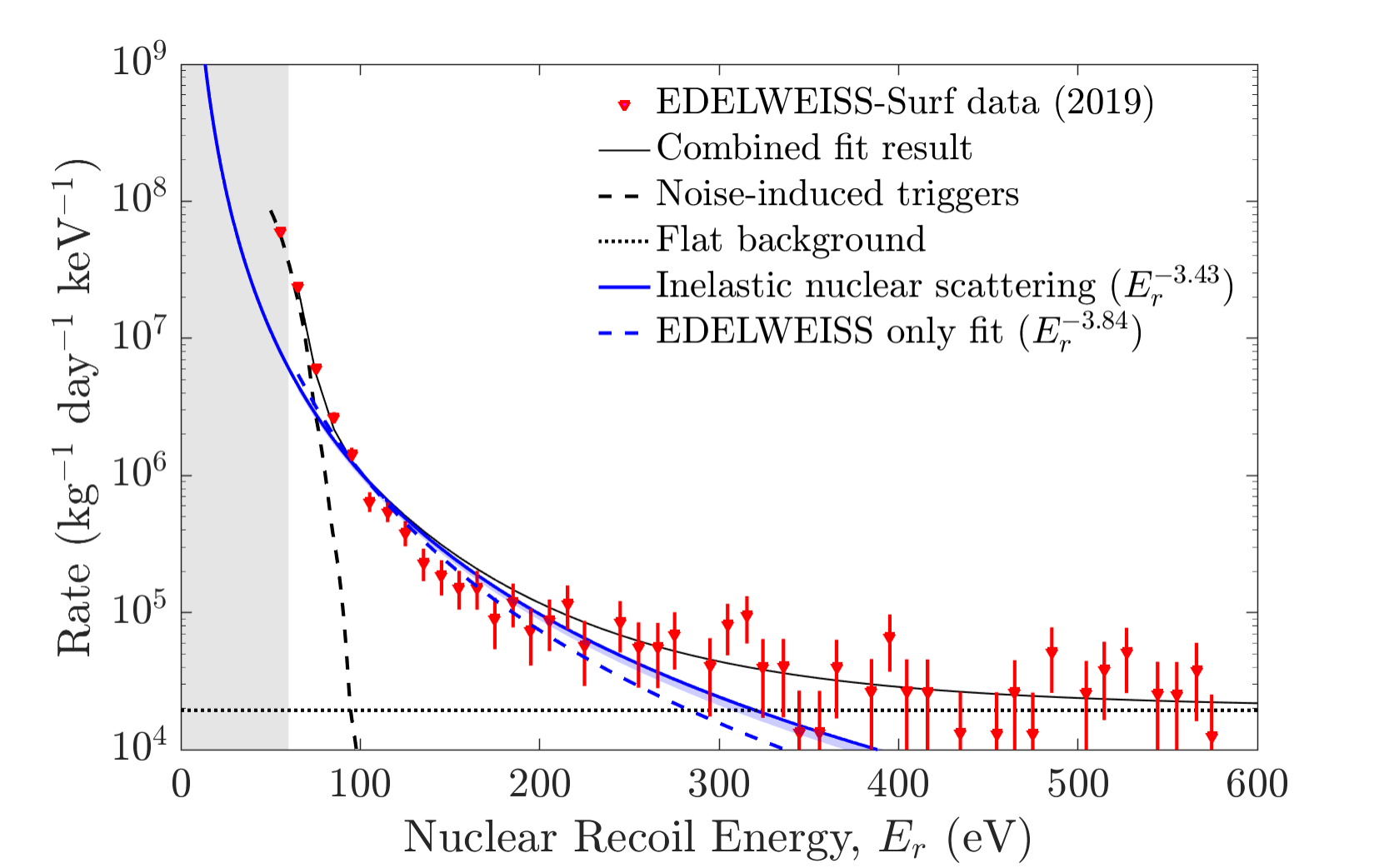}
    \caption{The efficiency-corrected calorimetric rates (red points) from the silicon SuperCDMS CPD~\cite{alkhatib2020light} (left) and germanium EDELWEISS-Surf~\cite{EdelweissWIMP} (right) detectors with statistical error bars are plotted against a global three-component fit (thin black line, Eq.~(\ref{eq:rate})) consisting of noise triggers above threshold (black dashed, see text for more details), a one-parameter flat component representing standard radiogenic backgrounds (black dotted), and excess events which are fit to a $E_r^{-3.43}$ power law dependence (solid thick blue). For the excess component, which is considered for this work as arising from inelastic nuclear scattering, a 1$\sigma$ uncertainty band from the combined fit is also shown (shaded blue), along with the fit to each dataset separately (dashed blue). In the combined fit, the two datasets are fit simultaneously, with separate noise trigger and flat background components for each detector but a common power law index. The detector thresholds are represented by the grey shaded regions at low energy. 
    }
    \label{fig:excesses}
\end{figure*}

This paper is organized as follows. In Sec.~\ref{sec:Analysis}, we review the recent progress in understanding the persistent excesses in low-threshold detectors, and perform a combined fit to the SuperCDMS CPD and Edelweiss-Surf excesses, demonstrating an intriguing consistency in spectral index and normalization which is suggestive of a possible DM interpretation. In Sec.~\ref{sec:DM}, we use a phenomenological model of the detector response, parametrized by the dynamic structure factor, to attempt to explain both excesses in the context of inelastic DM-nuclear scattering. We find that such an interpretation is inconsistent even allowing for exotic structure factors, largely due to the fact that the allowed region for the silicon excess rate requires dark matter masses small enough that they have insufficient kinetic energy to yield the measured germanium rate at higher energies. In Sec.~\ref{sec:KnownParticles}, we argue that the excess is also inconsistent with nuclear scattering from known particle sources, namely cosmic-ray neutrons, photons, and neutrinos, as well as secondary interactions. We conclude in Sec.~\ref{sec:Conclusions} with our summary of this puzzling situation: the calorimetric excesses remain robustly mysterious.

\section{Combined analysis of semiconductor excesses}
\label{sec:Analysis}

We noted in Ref.~\cite{plasmonKBKK} that there was significant discrepancy at the time among the excesses in the silicon ionization detectors SuperCDMS~HVeV, DAMIC at SNOLAB, and SENSEI~\cite{damicER_2019,Abramoff_2019,Agnese_2018}, each of which observed different single-electron dark rates. 
SENSEI has since released new results~\cite{barak2020sensei} from a detector operated with shallow 225~m.w.e.\ overburden  that reduced their measured single(multiple)-electron dark rate to 5(0.05)~Hz/kg, consistent with the DAMIC single-electron dark rate of 7~Hz/kg~\cite{damicER_2019} despite the increased shielding and 6000~m.w.e. overburden at SNOLAB. 
This resolved the initial tension mentioned in Ref.~\cite{plasmonKBKK} and indicated some unrelated origin for the single(multiple)-electron dark rate background in the SuperCDMS~HVeV detector of 1700(13)~Hz/kg~\cite{Agnese_2018,cdmsHV_2020}. Moreover, recent work~\cite{Du:2020ldo} has demonstrated consistency between some of these dark rates and secondary background processes, such as from Cherenkov emission, indicating a potential radiogenic contribution to these backgrounds.  

Thus, since there has been much progress toward understanding the excesses in ionization detectors, we now focus exclusively on a common interpretation of semiconductor calorimetric excesses, which remain mysterious. 
The SuperCDMS CPD~\cite{alkhatib2020light} excess in silicon is analogous to the earlier EDELWEISS-Surf measurement in germanium \cite{EdelweissWIMP} in that it measures the total recoil energy deposited in the detector $E_r$, regardless of the distribution of the primary event energy into heat or charge (less any persistent defect energies which are on the order of 4~eV per defect~\cite{jiang2018theoretical} and are neglected in this analysis). 
Both detectors are also notably operated on the surface with minimal shielding. 
Whereas in Ref.~\cite{plasmonKBKK}, we focused primarily on qualitative arguments to motivate further interest in these excesses, here we perform a more quantitative analysis of the SuperCDMS CPD~\cite{alkhatib2020light} and EDELWEISS-Surf~\cite{EdelweissWIMP} excess rates. 

The SuperCDMS CPD result is of particular interest because its threshold (25~eV) is considerably lower than that of EDELWEISS-Surf (60~eV). 
Both detectors measure an approximately exponential background near threshold which is likely from noise triggers that are not removed by the analysis cuts. 
In the case of EDELWEISS-Surf, a model for these noise-induced triggers was published in Ref.~\cite{EdelweissWIMP} and has been incorporated into this analysis directly, with no free parameters. 
For SuperCDMS CPD, these triggers are likely coming from environmental noise, and thus do strictly follow an exponential in energy. 
At higher energies, both detectors are limited by ``flat" radiogenic backgrounds (e.g., Compton scattering~\cite{Ramanathan:2017dfn}) on the order of $10^5$~cts~kg$^{-1}$~day$^{-1}$~keV$^{-1}$, as is to be expected for detectors operating on the surface. 
However, between these two distinct features, both detectors observe a statistically significant excess of events.
EDELWEISS explored in great detail the possibility that these excess events in germanium come from elastic or Migdal\footnote{Ref.~\cite{EdelweissWIMP} uses the isolated atom formalism~\cite{Ibe:2017yqa} to calculate these rates in germanium, which neglects important collective effects~\cite{tongyan_2020}; that said, the isolated atom approach was the only calculation in the literature at the time of publication.} scattering of DM particles and found that neither gives a good spectral match to the data~\cite{EdelweissWIMP}. 

We simultaneously fit the digitized SuperCDMS CPD and EDELWEISS-Surf data to the following model: a flat background $D_i$ for each detector (1 parameter each), a model for noise triggers leaking above threshold $f(E_r)$ (consisting of a 2-parameter exponential for SuperCDMS CPD and a 0-parameter model for EDELWEISS-Surf taken directly from Ref.~\cite{EdelweissWIMP} and scaled by signal efficiency to compare with data), and a power law component in recoil energy for the excess (with independent normalization for each detector and a common power law index for both, 3 parameters total):
\begin{equation}\label{eq:rate}
    \frac{dR_i}{dE_r} = (C \kappa^2)_i E_r^{-\alpha} +  D_i + f_i(E_r),
\end{equation}
where $i = {\rm Si}, \ {\rm Ge}$. We write the normalization of the excess in the suggestive form $(C \kappa^2)$ because the DM model we present in Sec.~\ref{sec:DM} will contain an overall normalization of the DM-nucleon cross section proportional to $C$ and a detector-dependent factor $\kappa^2$ which will be $A^2$ or $Z^2$ for DM which couples to nucleons or protons, respectively (here $A$ is the mass number and $Z$ is the atomic number of the target). Under the assumption of a common origin between the two detectors, the common normalization $C$ cancels yielding only a detector-dependent ratio $\kappa_{\rm Ge}^2/\kappa_{\rm Si}^2$.
This joint fit provides a best fit power law index of $\alpha = 3.43^{+0.11}_{-0.06}$ and a Ge-to-Si normalization ratio of $\kappa^2_{\rm Ge}/\kappa^2_{\rm Si} = 7.2 ^{+0.7}_{-0.6}$.  
The results of this fit are shown in Fig.~\ref{fig:excesses} and presented in Table~\ref{tab:fitparams}.\footnote{A similar analysis was recently performed on excess rates in sapphire~\cite{Heikinheimo:2021syx}, which focused instead on the potential creation of defects through an exotic power-law nuclear scattering channel.} We also perform individual fits to each detector spectrum, by allowing a power law index $\alpha_i$ which differs for each dataset. Notably, the fits when including the EDELWEISS-Surf data are worse, which would be significantly improved by adding additional fit parameters to capture the systematic uncertainties present in the high-side tail of the noise-induced trigger model for that data.

\begin{table*}[!t]
\begin{center}
\begin{tabular}{l@{\hskip 0.05in} | @{\hskip 0.1in}c@{\hskip 0.15in}c@{\hskip 0.1in} c@{\hskip 0.2in} c@{\hskip 0.15in} c@{\hskip 0.15in} c }
\hline \hline
\rule{0pt}{2.5ex} & $C \kappa^2_{\rm Si}$ & $C \kappa^2_{\rm Ge}$ & \multirow{2}{*}{$\alpha$} & \multirow{2}{*}{$\kappa^2_{\rm Ge}/\kappa^2_{\rm Si}$} & \multirow{2}{*}{$\chi^2/d.o.f.$} & \multirow{2}{*}{$p$-value} \\
& $[(\text{kg day})^{-1} \text{keV}^{\alpha-1} ]$ & $[(\text{kg day})^{-1} \text{keV}^{\alpha-1} ]$ & & & \\
\hline
\rule{0pt}{2.5ex}Combined fit (Fig.~\ref{fig:excesses}) & $10.8^{+5.6}_{-2.6} \times 10^{11}$ & $7.9^{+5.0}_{-2.3} \times 10^{12}$ & $3.43^{+0.11}_{-0.06}$ & $7.2^{+0.7}_{-0.6}$ & 227.2/174 & $0.004$  \\
\rule{0pt}{2.5ex}SuperCDMS CPD~\cite{alkhatib2020light} only & $4.1 \pm 1.5 \times 10^{11}$ & n/a & $3.18^{+0.10}_{-0.09}$ & n/a & 121.5/121 & $0.471$ \\
\rule{0pt}{2.5ex}EDELWEISS-Surf~\cite{EdelweissWIMP} only & n/a & $51.4^{+44.2}_{-25.9} \times 10^{12}$ & $3.84^{+0.13}_{-0.15}$ & n/a & 101.3/52 & $5 \times 10^{-5}$ \\
\hline \hline
\end{tabular}
\caption{\label{tab:fitparams}Values for the fit parameters of Eq.~\eqref{eq:rate} for each of the cases considered in the text. 
The $p$-values are calculated based on the listed Pearson $\chi^2$ values and degrees of freedom (number of bins minus parameters in each fit). 
Here, larger $p$-values indicate a better fit. 
The poor (but notably non-zero) $p$-value when including the EDELWEISS-Surf data is driven by the bin just below 100 eV, which is pulling the power law index up and is maximally sensitive to systematics in the (fixed) noise-induced trigger model, which are not included in the fit.  
Of particular note, all three fits give consistent power law indices $\alpha$, which is positively correlated with both the overall normalization $C$ and the ratio of the rates $\kappa^2_{\rm Ge}/\kappa^2_{\rm Si}$. 
}
\end{center}
\end{table*}

The fact that these two independent datasets from different collaborations with different sensor technologies and different target materials (but comparable mK temperatures) measure an excess of events at low energies following compatible power laws is by itself interesting.\footnote{\label{foot:Migdal}Curiously, the prediction of the Migdal effect from Ref.~\cite{Knapen:2020aky} is a power law with index $\alpha=4$ at these energies, but the total rate is inconsistent with the interpretation of these events as coming from DM scattering through the Migdal effect, and furthermore the large energy deposit to the electronic system would likely result in excess ionization yields which are not seen in the ionization detectors at these energies~\cite{edelweissHV,damic2020}.} Independent of their respective rates, this common power law is potentially indicative of a similar (or even identical) physical process as the origin of these events in each detector. 
Notably, the excess rates in each of these detectors can also be individually fit to an exponential, rather than a power law; however, the different energy ranges of the two excess rates exclude the possibility of a common exponential decay constant. 
In addition to the common power law index $\alpha \approx 3.4$, the fact that the ratio of the $\kappa^2$ is consistent with the ratio of the square of the mass numbers $A$ of the two targets, $(74/28)^2 = 7.0$, is intriguing given that the standard benchmark model of spin-independent DM-nuclear scattering
scales precisely in this fashion. 

\section{Dark matter interpretation through exotic structure factors}
\label{sec:DM}
To see whether the observed excess might be consistent with a DM interpretation yielding a power-law energy spectrum $dR/d\omega \propto \omega^{-3.4}$,  here we consider a generic formalism for calculating the rates for DM-nuclear scattering on solid-state targets using an empirical parametrization of the dynamic structure factor that allows for physically-allowed, but nontrivial, collective effects without necessarily requiring a microscopic interpretation. In what follows, we will refer to the energy deposited by the DM as $\omega$ rather than $E_r$, to emphasize the fact that collective effects may play a role and that we are not dealing with the elastic recoil energy of a single isolated nucleus. 

Assuming nothing about the DM-detector system other than the validity of the Born approximation, the DM-nuclear scattering rate may be expressed in terms of the dynamic structure factor~\cite{Trickle:2019nya}, which encapsulates the response of the target to a perturbation of the ion density,
\begin{equation}
S(\mathbf{q}, \omega)= 2\pi \sum_{\beta}\left|\left\langle\Psi_{\beta}\left|n_{\mathbf{q}}\right| \Psi_{0}\right\rangle\right|^{2} \delta\left(\omega-\omega_{\beta}\right).
\end{equation}
Here, $|\Psi_0 \rangle$ is the ground state of the system, $|\Psi_\beta \rangle$ runs over all final states, and $n_{\bf q}$ is the density operator in Fourier space, 
\begin{equation}
n_{\mathbf{q}}=\frac{1}{\sqrt{V}} \sum_{j} e^{i \mathbf{q} \cdot \mathbf{r}_{j}},
\end{equation}
where $\mathbf{r}_{j}$ are the positions of all the nuclei in the target and $V$ is the detector volume. 
When the dynamic structure factor is isotropic, $S(\vecq, \omega) = S(q, \omega)$, the differential DM scattering rate per unit target mass can be obtained from the structure factor as \cite{Trickle:2019nya}
\begin{equation}
\label{eq:Sspectrum}
\frac{d R}{d \omega} =\frac{\rho_{\chi}}{m_{\chi}} \frac{\kappa^2\bar{\sigma}_{n}}{2 \mu_{\chi n}^{2}} \frac{1}{2 \pi \rho_{T}} \int dq \, q \, S(q, \omega) \eta\left(v_{\min }\right) ,
\end{equation}
where $\rho_\chi =0.3~\text{GeV}\,\text{cm}^{-3} $ is the local DM density; $\bar{\sigma}_n$ is the fiducial DM-proton or DM-nucleon cross section; $\kappa^2 = Z^2$ or $A^2$ depending on whether the DM couples to protons or all nucleons, respectively; $\rho_T = m_N n_0$ is the target mass density; $\eta(v_{min})$ is the DM mean inverse speed; and $v_{\rm min}$ is the minimum DM speed required to deposit energy $\omega$,
\begin{equation}
\label{vmin1}
v_{\min }(q, \omega)=\frac{\omega}{q}+\frac{q}{2 m_{\chi}}.
\end{equation}
We have assumed a heavy mediator such that the cross section is independent of $q$ (i.e. $F_{\rm DM}(q) = 1$), both for simplicity and to more easily make contact with experimental limits making the same assumption~\cite{lewin}. Similarly, in the DM mass range we will be interested in, there is insufficient momentum to probe nuclear substructure and so we also set the nuclear form factor to unity. It is clear from Eq.~(\ref{eq:Sspectrum}) that a choice of $S(q,\omega)$ fully determines the spectral shape of the differential scattering rate, given a choice of DM velocity distribution. The integrated rate requires further input from the DM interaction strength, parametrized by $\kappa^2 \overline{\sigma}_n$.

A first-principles computation of the structure factor is possible in specific simplified models, including treating $|\Psi_0 \rangle$ and $|\Psi_\beta \rangle$ as single-particle harmonic oscillator states or plane waves \cite{Kahn:2020fef,Knapen:2020aky}. However, when considering the energy deposit to the scattered nucleus alone (in contrast to the $\omega^{-4}$ electronic energy spectrum noted in Footnote~\ref{foot:Migdal}), such a model either yields the ordinary flat spectrum of elastic scattering when $q \simeq \sqrt{2 m_N \omega}$, or a steeply-falling spectrum $dR/d\omega
\propto \exp[-\omega^2 m_N/(q_{\rm max}^2 \omega_0) ]$ when $q \ll \sqrt{2 m_N \omega}$, where $\omega_0 \simeq 60 \ {\rm meV}$ is the optical phonon energy in Si or Ge and $q_{\rm max} \simeq 2 m_\chi v$ is the maximum momentum transfer. Both of these spectral shapes are clearly inconsistent with the data.

\begin{figure*}[!t]
    \centering
    \includegraphics[width=0.49\textwidth, trim=0 0 0 0, clip=true]{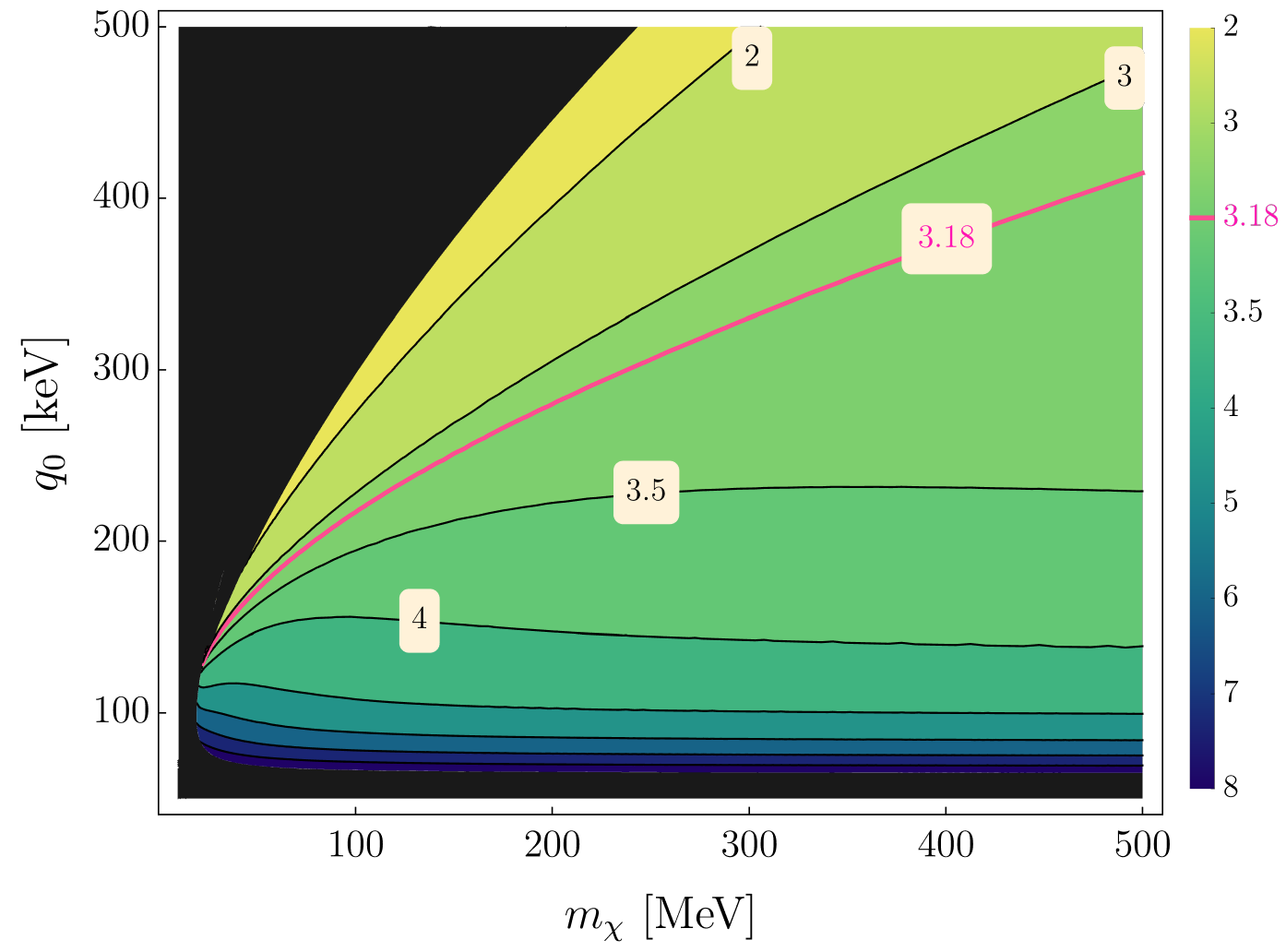}
    \includegraphics[width=0.49\textwidth, trim=0 0 0 0, clip=true]{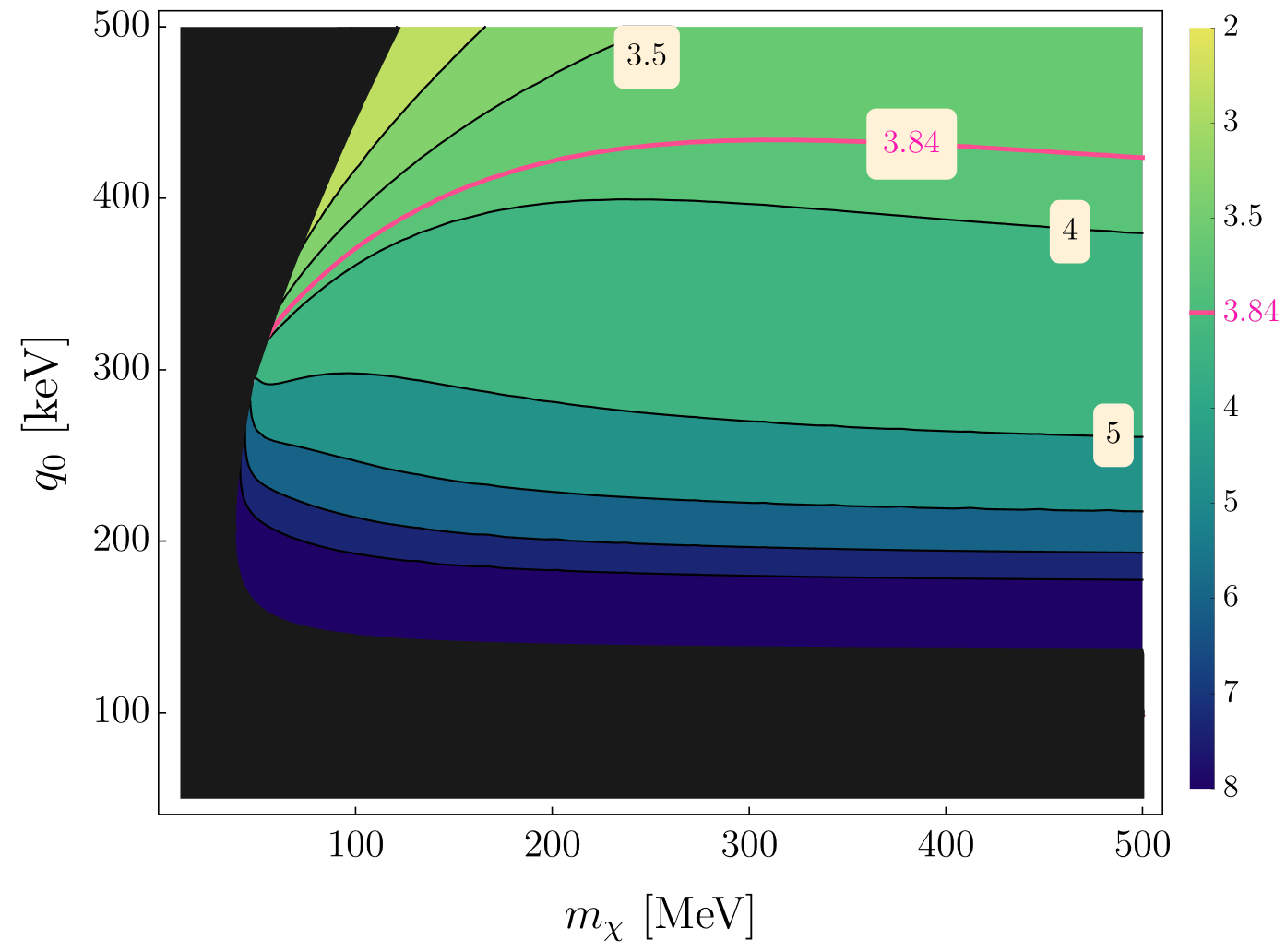}
    \includegraphics[width=0.49\textwidth, trim=0 0 0 0, clip=true]{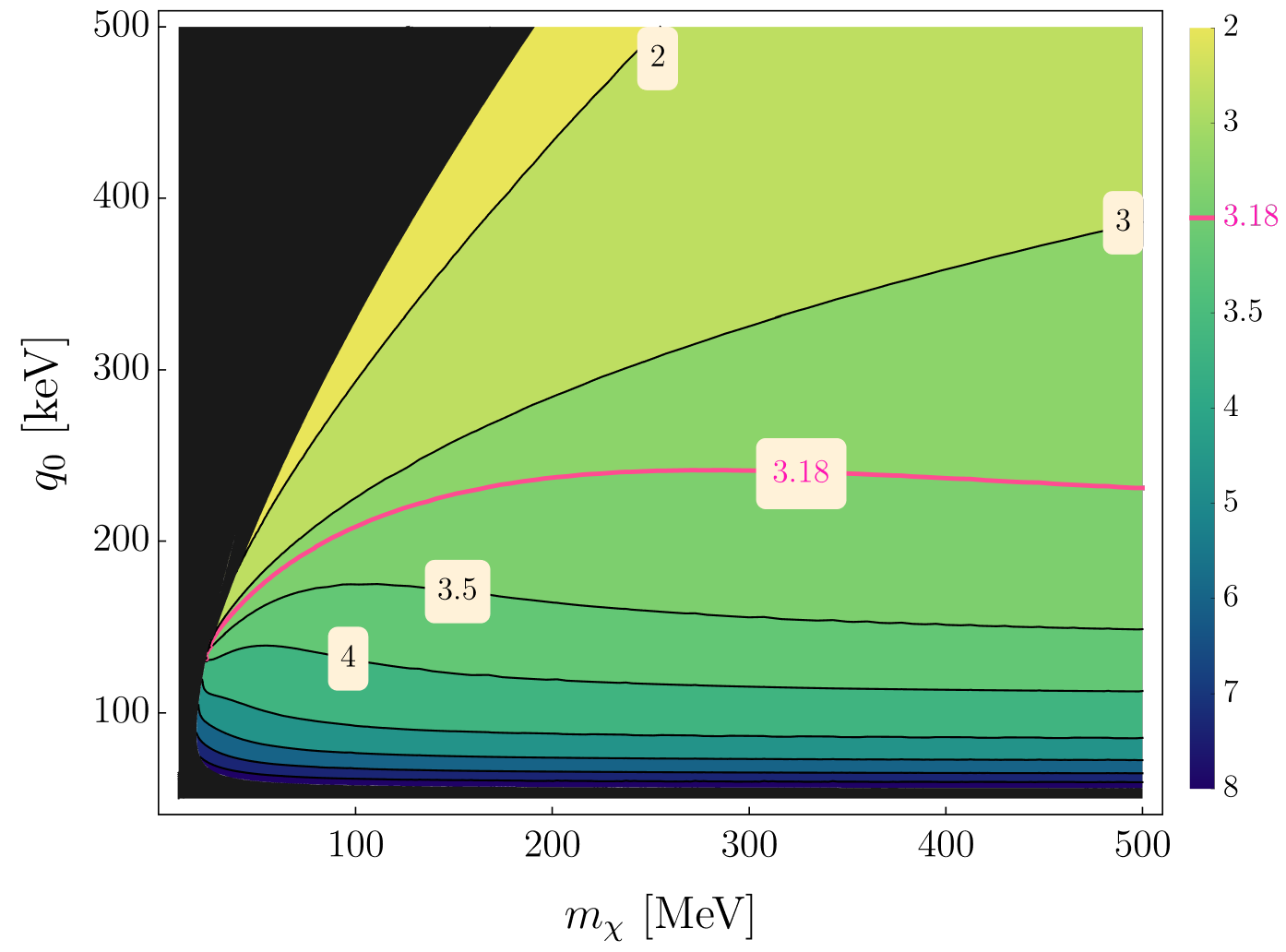}
    \includegraphics[width=0.49\textwidth, trim=0 0 0 0, clip=true]{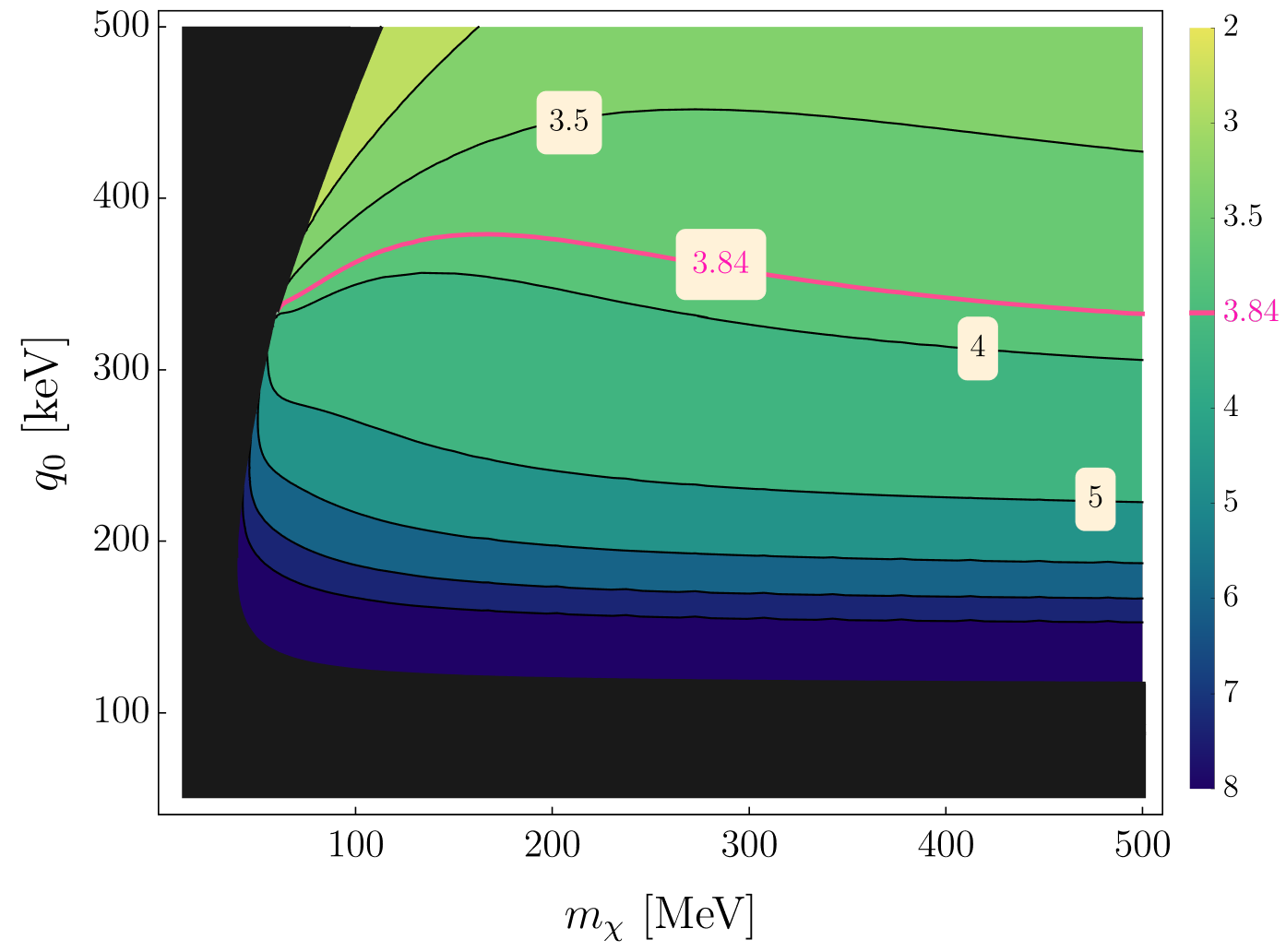}
    \caption{Contours of the power-law index $\alpha$ dependence of $dR/d\omega$ in silicon (left) and germanium (right) as a function of $q_0$ and $m_\chi$ with $S(q) = 0.015(q/q_0)^n$ for $n=5$ (top) and $n=6$ (bottom), from Eq.~(\ref{eq:approxspectrum}). The contours in each
    panel represent values of $m_\chi$ and $q_0$ which consistently yield a power law with the labeled value of $\alpha$, as shown in Eq.~(\ref{eq:approxspectrum}). The black shaded regions yield zero events at $\omega = 75$ eV for Si and at $\omega = 150$ eV for Ge (here, $\omega \equiv E_r$) and/or a non-monotonic spectrum, either of which is inconsistent with the data. The best-fit contours for each, $\alpha = 3.18$ for Si and $\alpha = 3.84$ for Ge, are shown in magenta.
    }
    \label{fig:spectralindexes}
\end{figure*}

To attempt to reproduce the observed power-law spectrum, we first suppose that the structure factor is dominated by a single-quasiparticle excitation, representing a single scattered nucleus interacting with the surrounding electron density. In this case, the dynamic structure factor may be parametrized as
\begin{equation}
\label{eq:Sqw}
S(\mathbf{q}, \omega)=2 \pi n_0 S(\mathbf{q}) \delta\left(\omega-\frac{q^{2}}{2 m_{N} S(\mathbf{q})}\right).
\end{equation}
The function $S(\mathbf{q})$ is known as the static structure factor and parametrizes departures from the free-particle dispersion relation: if $S(\mathbf{q}) = 1$, the dispersion relation is $\omega = q^2/(2m_N)$ as expected for elastic nuclear recoil, but static structure factors which differ from unity permit different dispersions. Furthermore, for \emph{any} choice of $S(\mathbf{q})$, $S(\mathbf{q}, \omega)$ in Eq.~(\ref{eq:Sqw}) automatically satisfies the ``$f$-sum rule''
\begin{equation}
\label{eq:fsum}
    \int_0^\infty \frac{d\omega}{2\pi} \, \omega S(\mathbf{q}, \omega) = \frac{q^2}{2m_N} n_0,
\end{equation}
which is a consistency condition on physically-realizable dynamical structure factors imposed by causality and conservation of mass.

We now make an ansatz for the form of $S(\vecq)$ designed specifically to yield the desired power-law spectrum. Suppose that the static structure factor is isotropic and itself follows a power law,
\begin{equation}
   S(q) \approx A_q (q/q_0)^n \qquad (A_q = 0.015),
\end{equation}
over a limited range of $q$ around a fiducial momentum value $q_0$. 
The prefactor $A_q$ may of course be absorbed into $q_0$, but is explicitly separated here to better illustrate typical kinematics: if $m_\chi = 200 \ {\rm MeV}$, its typical momentum is $q \sim 200 \ {\rm keV}$, and in order for $S(\mathbf{q},\omega)$ to have support at $\omega = 50 \ {\rm eV}$ and $q = 200 \ {\rm keV}$, we must have $S(q) = q^2/(2m_N \omega) = 0.015$ for $m_N = 26 \ {\rm GeV}$ in silicon. Indeed, the fact that $A_q \ll 1$ (so $S(q) \ll 1$ for $q$ near $q_0$) reflects the highly inelastic nature of the scattering interpretation of the excess: much more energy is deposited for a given momentum transfer than would be expected from elastic scattering. The free parameters in this model are thus the momentum scale $q_0$ and the power law index $n$. 
We emphatically do \emph{not} attempt any microscopic explanation of such a structure factor, but simply note that the (uncalibrated) energy regime we are concerned with here is just above the typical displacement energy in Si and Ge required to remove a nucleus from its lattice site, and thus we might expect qualitatively different behavior than in the single-phonon or high-energy ballistic recoil regimes, perhaps due to binding potential effects which distort the outgoing wave function, and/or interactions of the charged recoiling ion with the electron system.

Plugging in the power law ansatz for $S(\mathbf{q})$ into Eq.~(\ref{eq:Sspectrum}), and rearranging to emphasize the similarities to Eq.~(\ref{eq:rate}), yields
\begin{equation}
\label{eq:spectrum}
    \frac{dR}{d\omega} = \frac{\rho_{\chi}}{m_{\chi}} \frac{\kappa^2\bar{\sigma}_{n}}{2 \mu_{\chi n}^{2}} \frac{2A_q^2}{|2-n|}
    \left(\frac{ F_n(\omega)}{q_0}\right)^{2n}  \, \eta\left[v_{\min }(\omega)\right] ,
\end{equation}
where we have defined the dimension-1 quantity
\begin{equation}
F_n(\omega) \equiv \left(\frac{2m_N A_q \omega}{q_0^n}\right)^{\frac{1}{2-n}}~,
\end{equation}
such that the minimum velocity becomes  
\begin{equation}
    v_{\rm min}(\omega) = \frac{\omega}{
    F_n(\omega)}
     + \frac{
    F_n(\omega)}{2m_\chi}.
\end{equation}
Note that $F_0(\omega) = q$ for the elastic case with $A_q = 1$. If $v_{\rm min}(\omega)$ is independent of $\omega$ (which is approximately true for sufficiently large $m_\chi$), then Eq.~(\ref{eq:spectrum}) reduces to
\begin{equation}
\label{eq:approxspectrum}
    \frac{dR}{d\omega} \approx (C \kappa)^2 \omega^{-\alpha}~,
\end{equation}
and the spectrum is (by construction) exactly a power law with $\alpha = 2n/(n-2)$. Including the effects of $\eta(v_{\rm min})$ will distort the spectrum for smaller $m_\chi$, since less kinetic energy and less momentum are available for scattering, as well as for small $q_0$ which pushes the scattering to the high-velocity tail. Therefore, Eq.~(\ref{eq:approxspectrum}) is approximate and results from fitting the full spectrum to a power law. In particular, taking $n = 5~(n =6)$ yields $dR/d\omega \propto \omega^{-3.3}~(dR/d\omega \propto \omega^{-3})$ up to velocity-suppression effects (which would begin to increase the effective $\alpha$, rapidly in the case of Ge).

Figure~\ref{fig:spectralindexes} shows the results of fitting the Si and Ge spectra $dR/d\omega$ with a power law, as a function of $q_0$ and $m_\chi$ with $S(q)= 0.015 (q/q_0)^n$ for $n=5,6$. We see that achieving the index of $\alpha \approx 3.4$ preferred by the data is allowed in Si for a wide range of values for both $m_\chi$ and $q_0$. The higher energies of the excess in Ge make the same power law fit difficult because of the effects of velocity suppression, yielding a much narrower parameter space which does not overlap in $q_0$ with the Si best-fit contours except at the largest DM masses. At this stage, the difficulty of fitting both spectra simultaneously is clear, at least assuming that Si and Ge have comparable structure factors. 

\begin{figure}[!t]
\includegraphics[width=\columnwidth]{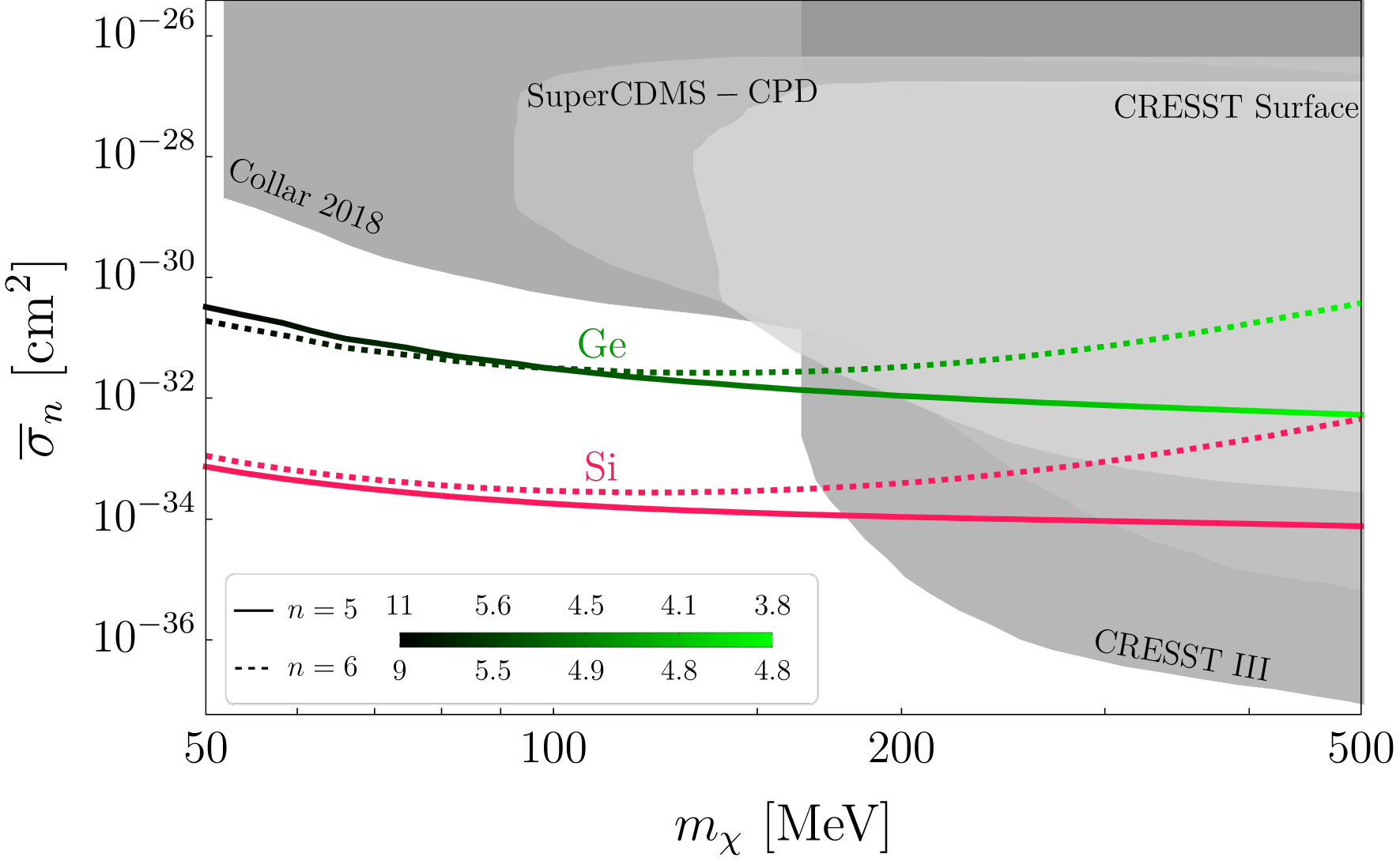}
\caption{Parameter space for a DM interpretation of the excess rates in SuperCDMS CPD and EDELWEISS-Surf. The solid (dashed) magenta contour corresponds to an integrated rate of 0.6 Hz/kg in SuperCDMS CPD for $\omega \in [40, 100]$ eV, $S(q) = A_q(q/q_0)^n$ for $n =5$~($n=6$) with $A_q = 0.015$, and $q_0$ chosen along the best-fit power law contour $\alpha = 3.18$ from Fig.~\ref{fig:spectralindexes}. The green contour shows the same structure factor applied to EDELWEISS data, normalized to a total rate of 1.3 Hz/kg for $\omega \in [100, 250]$ eV; the color gradient indicates the power law index, which is everywhere steeper than the best-fit $\alpha = 3.84$. Both contours correspond to a DM-nucleon interaction with $\kappa^2 = A^2$. The mismatched power law indices and DM-nucleon cross sections between the two experiments indicate the tension in a DM interpretation. 
Also included are elastic DM-nucleon scattering limits from CRESST~\cite{Abdelhameed_2019,CRESST:2017ues}, SuperCDMS CPD~\cite{alkhatib2020light}, and EJ-301~\cite{collarEJ}.
}\label{fig:CS}
\end{figure}

Once the power-law dependence of $S(q)$ is fixed, the normalization of the spectrum is also fixed up to the overall scaling by $\bar{\sigma}_n$. For the same $n = 5~(n = 6)$ dependence of $S(q)$ and taking $\kappa^2 = A^2$, the magenta solid (dashed) line in Fig.~\ref{fig:CS} shows the preferred region of $\overline{\sigma}_n$  and $m_\chi$ which yields an integrated rate of 0.6 Hz/kg in silicon for $\omega \in [40, 100]$~eV, and also yields a spectrum with the individual best-fit power law index $\alpha =  3.18$ for the CPD data in that energy range. Points on the magenta curves in Fig.~\ref{fig:CS} correspond to taking parameters along the corresponding $(m_\chi, q_0)$ magenta contours in Fig.~\ref{fig:spectralindexes}. We see that heavier dark matter masses $m_\chi \gtrsim 170 \ {\rm MeV}$ are robustly excluded by CRESST-III~\cite{Abdelhameed_2019}, but intriguingly, the preferred region for lower DM masses $m_\chi \simeq 100 \ {\rm MeV}$ is not excluded by any nuclear scattering experiment. Note that including an additional elastic term in the structure factor which has support at the same values of $\omega$ amounts to taking $S(\vecq) = 1$ in a regime of $\vecq$ distinct from the one where the inelastic structure factor has support. Since the sum rule in Eq.~(\ref{eq:fsum}) fixes the normalization of the structure factor at all $\vecq$, such an elastic contribution would only serve to increase the rate, and therefore in principle this could push the preferred values of $\bar{\sigma}_n$ slightly lower. However, for $m_\chi \lesssim 400 \ {\rm MeV}$, DM with velocity below the lab-frame galactic escape velocity cannot yield an elastic nuclear recoil in Si with energy above 40 eV, so all contributions to the observed excess above the exponential noise trigger must come from the inelastic structure factor.

Taking the same structure factor parameters along the Si best-fit contours in Fig.~\ref{fig:spectralindexes}, the green line in Fig.~\ref{fig:CS} shows the region of $\overline{\sigma}_n$ and $m_\chi$ which yields a total rate of 1.3 Hz/kg in germanium for $\omega \in [100, 250]$~eV, with the gradient indicating the power law dependence of the spectrum. Not only is the cross section $\overline{\sigma}_n$ inconsistent with Si, but the Ge spectrum is everywhere too steep to match the best-fit value of $\alpha = 3.84$, except near $m_\chi \simeq 500 \ {\rm MeV}$ which is excluded by several other experiments. It is possible that the cross sections may be brought into agreement by widely differing values of $A_q$ (or equivalently $q_0$) between Si and Ge, but the fact that the allowed region for Si is restricted to $m_\chi \lesssim 170 \ {\rm MeV}$ means that only DM on the high-velocity tail of the DM distribution has enough kinetic energy to generate events in the 200 eV range, regardless of the structure factor. This will always serve to steepen the power law index beyond what is observed in the data and renders the simultaneous DM interpretation of the SuperCDMS CPD and EDELWEISS-Surf data highly implausible. 

Even attempting to explain one or the other of the excesses, rather than both, requires an extremely peculiar inelastic dispersion $\omega \propto q^{-3}$ which arises from $S(q) \propto q^5$. That said, systems with such a dispersion, where the energy of the excitation decreases with increasing momentum, are not unheard of; indeed, superfluid helium exhibits this phenomenology between the maxon and roton regions~\cite{donnelly1981specific}, as do plasmons in some transition metal dichalcogenides \cite{van2011effect}. Testing this explanation of the excess would require measuring the structure factor in semiconductors with neutron scattering, exactly as was done to determine the structure factor of helium, but with momentum transfers on the order of $q_0$ and energy deposits in the 40--100 eV energy range. 

\section{Ruling out known particle sources}
\label{sec:KnownParticles}
 Since even rather unusual condensed matter effects are unable to furnish a consistent DM interpretation, and given that both detectors were operated on the surface, we also examine the possibility that the excess is due to cosmic-ray (CR) neutron scattering. The CR neutron spectrum at ground level is very close to flat in $\ln E_n$, where $E_n$ is the CR neutron energy,
\be
\frac{d\Phi}{d \ln E_n} \approx \Phi_0,
\ee
varying only by $\mathcal{O}(1)$ factors over 10 orders of magnitude between $E_n = 10 \ {\rm meV}$ and $E_n = 100 \ {\rm MeV}$ \cite{IBMroof,SatoCR}. Here $\Phi_0 \approx 1 \times 10^{-3} \, {\rm Hz}/{\rm cm}^{-2}$ is the approximate CR neutron flux at sea level. This spectrum translates to a CR neutron velocity distribution $f(v) \propto 1/v^2$. In the case of elastic scattering parametrized by a neutron-nucleus cross section $\sigma_{nN}$, this leads to an energy spectrum
\be
\left. \frac{dR_{{\rm CR}n}}{d\omega}\right|_{\rm el.} = \frac{\Phi_0 \sigma_{nN}}{m_N} \omega^{-1},
\ee
which has the wrong power-law index to match the observed excess. Moreover, for Si, taking $\sigma_{nN} = 4\pi a^2$, where $a = 4.2 \ {\rm fm}$ is the neutron scattering length in Si, the total rate between 40 and 100 eV is $\simeq 0.05$ Hz/kg, a factor of 10 below the measured excess rate. In order to achieve a power-law spectrum $\omega^{-3.18}$ and an integrated rate of 1 Hz/kg, one would have to postulate a neutron energy spectrum $d\Phi/dE_n \propto E_n^{-3.18}$ with a total neutron flux more than 5 orders of magnitude larger than $\Phi_0$ since the different neutron spectrum implies a different normalization for neutrons of the appropriate energy. Even if we further speculate a peculiar inelastic dispersion from a nontrivial structure factor $S(q)$, which could reconcile the spectral index of the excess with the observed log-flat CR neutron spectrum, the large observed rate would still require a substantial additional flux of neutrons that is not observed.

Similar reasoning can rule out nuclear recoils induced from known fluxes of either incident neutrinos or photons. While exotic structure factors of the kind considered in Sec.~\ref{sec:DM} can change the spectral shape, the overall normalization is still driven by the total cross section for photons or neutrinos scattering off an individual nucleus,
\begin{align}  
    \sigma_{\nu N} & \approx \frac{G_F^2}{4\pi}Q_W^2 m_N \omega_{\rm max} = 2.3 \times 10^{-42}  \ {\rm cm}^2 \\
    \sigma_{\gamma N} & \approx \frac{8\pi}{3}\frac{Z^4 \alpha^2}{m_N^2} = 1.0 \times 10^{-29} \ {\rm cm}^2,
\end{align}
where $Q_W = N - Z(1 - 4 \sin^2 \theta_W)$ is the weak charge of a nucleus with $N$ neutrons and $Z$ protons, and we have given the numerical values for silicon $(\omega_{\rm max} = 140 \ {\rm eV})$. For the photon cross section, we have assumed coherent Thomson scattering from X-ray or gamma-ray photons with $E_\gamma \ll m_N$, which dominates over other photonuclear processes such as Delbr\"{u}ck scattering and resonant scattering \cite{Berghaus:2021wrp}. For the neutrino cross section we have assumed coherent scattering with $\omega \ll E_\nu$ and sufficiently low momentum transfer that the nuclear form factor is approximately unity, a reasonable approximation even for elastic scattering in the energy range relevant for the excesses. The total rate per unit mass for incident particle $X$ is $R = \Phi_X N_T \sigma_{XN}$, where $N_T \approx 2 \times 10^{25}/{\rm kg}$ is the number density of nuclei in Si, $\Phi_X$ is the flux of particles in question, and $\sigma_{XN}$ its cross section with nuclei.

For neutrinos, the total flux at the surface is dominated by keV solar neutrinos, $\Phi_\nu \approx 5 \times 10^{10}$ Hz/cm$^2$ \cite{Baxter:2021pqo}. The largest the rate can possibly be is if all of these neutrinos contributed to scattering (of course, this would also require a highly inelastic structure factor), in which case the total rate would be at most
\be
R_\nu \approx 3 \times 10^{-6}  \ {\rm Hz/kg},
\ee
clearly ruling out solar neutrinos.

We can estimate the total photon flux from the measured Compton rate in the SuperCDMS CPD detector, which is approximately $10^5$ events kg$^{-1}$ day$^{-1}$ keV$^{-1}$ at low energies. If we conservatively assume this rate is flat out to 1~MeV and integrate over this full range, we get an integrated rate of $\sim 10^3 \ {\rm Hz/kg}$. The Compton cross section per electron is approximately $\sigma_{\gamma e} = (8\pi/3) \alpha^2/m_e^2$, so the ratio of nuclear Thomson to Compton cross sections (accounting for the $Z$ electrons per nucleus) is $\sigma_{\gamma N}/\sigma_{\gamma e} = Z^3 m_e^2/m_N^2 = 2 \times 10^{-4}$ in Si. An upper bound on the total nuclear scattering rate from these photons can be obtained by rescaling the measured Compton rate, yielding for silicon
\be
R_\gamma \lesssim 0.2 \ {\rm Hz/kg},
\ee
which is close to the observed excess rate. However, the maximum elastic recoil energy for $E_\gamma = 1 \ {\rm MeV}$ in Si is 77 eV, so to explain the excess with elastic scattering, \emph{all} of the photons contributing to the Compton rate must have energies around or above 1 MeV, and the photon spectrum must be a power law with the correct index. 
Including inelastic structure factors will not improve the situation. In order to take advantage of the large number of photons at low energies, we would need $S(q) \ll 1$,  which would suppress the total rate well below that of the observed excess. 
Furthermore, the flat background rate of the EDELWEISS-Surf detector is actually \emph{lower} than in the SuperCDMS CPD data, indicating that the excess rates scale inversely to the ratio of radiogenic ionization backgrounds; this fact additionally disfavors a traditional radiogenic origin of these rates. 

An alternative possibility is that secondary interactions in material surrounding the detector may contribute to this low energy background, such as Cherenkov emission, decay of metastable states, or thermal events coupling into the detector via clamps. In the case of Cherenkov emission, this possibility was excluded in the analysis presented in Ref.~\cite{Du:2020ldo}. 
For the second case, we would expect a Poisson distribution of events in energy, which does not resemble the power law we have shown in this work; even a Poisson distribution with small mean would resemble an exponential with common decay constant between both detectors, which as argued in Sec.~\ref{sec:Analysis} is inconsistent with the distinct energy regimes of the two excesses. For both Cherenkov emission and metastable states, events would have to be modeled on a case-by-case basis as in Ref.~\cite{Du:2020ldo}, which involves enough free parameters that any analysis is fundamentally underdetermined, and thus while it is possible to create a power law in a limited regime, this explanation would be demonstrative but not fundamental proof of this mechanism. The thermal coupling scenario is largely ruled out for athermal detectors such as SuperCDMS CPD, as thermal events in surrounding materials can usually be rejected by pulse shape discrimination~\cite{pulseShape}.

In summary, all NR explanations for the measured excesses from fluxes of known particles seem rather implausible, and indeed all could easily be falsified with additional shielding in future runs of the experiments. However, this analysis also reveals the importance of developing new, lower energy neutron calibration methods which could be used to probe the low-energy, low-$q_0$ kinematic regimes considered in this analysis. The methods used in this paper are useful for excluding possible event origins based on allowable structure factors, but we stress the importance of measuring the features of this inelastic scattering regime.

\section{Conclusions}
\label{sec:Conclusions}
We have demonstrated that the SuperCDMS CPD and EDELWEISS-Surf excess rates can be modeled by a common power law of index $\alpha \approx 3.4$. Using a novel approach to quantitatively parametrize a physically-allowable dynamic structure factor which could yield such a power-law spectrum in the uncalibrated kinematic regime where inelastic effects may be expected, we argue that these two excess rates cannot be explained by a common origin involving inelastic nuclear recoil. In particular, the SuperCDMS CPD silicon data excludes the DM explanation for the EDELWEISS-Surf germanium excess, but itself could still be consistent with DM of mass $\lesssim 200$~MeV scattering through a highly-inelastic, novel nuclear recoil channel.  Moreover, the rates from both of these experiments are too high to be explained by nuclear scattering from any standard backgrounds, including neutrons, solar neutrinos, or photons. We thus conclude that these excesses are likely not due to a novel inelastic scattering process as originally proposed in Ref.~\cite{plasmonKBKK}, which bolsters the evidence for detector effects as a likely origin. That said, our analysis demonstrates the value of exploring compatibility between low-energy experimental excess rates in widely-varying detector environments, which can be a powerful tool for disentangling complicated new physics at these energies. \\

\emph{Acknowledgments.} None of the observations in this paper would be possible without the experimental results we cite, but also without private conversations with the collaborations responsible. 
We thus want to acknowledge (in alphabetical order) Ray Bunker, Alvaro Chavarria, Enectali Figueroa-Feliciano, Lauren Hsu, Paolo Privitera, Florian Reindl, and Belina von Krosigk. 
We thank Gordon Baym, Dan Hooper, Rocky Kolb, and Ben Safdi for useful conversations related to the content of this paper.
We are especially grateful to Julian Billiard, Juan Collar, Rouven Essig, Juan Estrada, Jules Gascon, Matt Pyle, Alan Robinson, and Felix Wagner 
for their feedback on early drafts of this analysis.
We thank the Gordon and Betty Moore Foundation and the American Physical Society for the support of the ``New Directions in Light Dark Matter'' workshop where the key idea for this work was conceived. 
We thank the EXCESS workshop organizers and participants for continuing the discussion of these experimental excess rates. 
P.A. acknowledges support from the Gordon and Betty Moore Foundation through EPiQS grant GBMF-9452. 
The work of Y.K. is supported in part by US Department of Energy grant DE-SC0015655. L.K.W. assisted in the theoretical modeling and editing of the paper, and was supported by the U.S. Department of Energy, Office of Science, Office of Basic Energy Sciences, Computational Materials Sciences program under Award Number DE-SC0020177. 
Fermilab is operated by Fermi Research Alliance, LLC, under Contract No. DE-AC02-07CH11359 with the US Department of Energy.  
This material is based upon work supported by the U.S. Department of Energy, Office of Science, National Quantum Information Science Research Centers, Quantum Science Center.
This work was supported in part by the Kavli Institute for Cosmological Physics at the University of Chicago through an endowment from the Kavli Foundation and its founder Fred Kavli.

\bibliography{dm}

\end{document}